\documentclass[reprint]{revtex4-1}
\usepackage{amsmath,color,float,graphicx}

\newcommand{\pe}{\text{Pe}}
\newcommand{\blu}[1]{#1}

\begin{document}

\title{
  Adsorption inhibition by swollen micelles may cause multistability in active droplets}
\author{Matvey Morozov}
\email{matvey.morozov@ulb.ac.be}
\affiliation{
  Nonlinear Physical Chemistry Unit, Facult{\'e} des Sciences, 
  Universit{\'e} libre de Bruxelles (ULB), CP231, 1050 Brussels, Belgium}

\begin{abstract}
Experiments indicate that microdroplets undergoing micellar solubilization in the bulk of surfactant solution may excite Marangoni flows and self-propel spontaneously. Surprisingly, self-propulsion emerges even when the critical micelle concentration is exceeded and the Marangoni effect should be saturated. To explain this, we propose a novel model of a dissolving active droplet that is based on two fundamental assumptions: (a)~products of the solubilization may inhibit surfactant adsorption; (b)~solubilization prevents the formation of a monolayer of surfactant molecules at the droplet interface. We use numerical simulations and asymptotic methods to demonstrate that our model indeed features spontaneous droplet self-propulsion. Our key finding is that in the case of axisymmetric flow and concentration fields, two qualitatively different types of droplet behavior may be stable for the same values of the physical parameters: steady self-propulsion and steady symmetric pumping. Although stability of these steady regimes is not guaranteed in the absence of axial symmetry, we argue that they will retain their respective stable manifolds in the phase space of a fully 3D problem.
\end{abstract}

\maketitle

%%%%%%%%%%%%%%%
% Introduction
%%%%%%%%%%%%%%%
\section{Introduction}
Chemically active microdrops submerged in the bulk of reagent solution may excite a flow in the surrounding fluid~\cite{Maass16}. For instance, hydrolysis of fatty acid precursors may enable spontaneous self-propulsion of microdroplets in the bulk of fatty acid solution~\cite{Hanczyc07}, while spontaneous motion and emulsification was observed in droplets producing amino acid-based surfactants~\cite{Nagasaka17}. Motility of microdroplets was also achieved in a wide class of systems featuring micellar solubilization in the case of both normal~\cite{Kruger16, Moerman17, Suga18} and inverse emulsions~\cite{Izri14}. In experiments, dissolving droplets may self-propel along a straight or chaotic trajectory~\cite{Izri14, Maass16, Moerman17, Suga18}, while droplets of nematic liquid crystal also exhibit helical self-propulsion regime~\cite{Kruger16, Suga18}. Multiple active drops ``feel'' each other’s presence and adjust their behavior: they may form ordered clusters~\cite{Thutupalli11, Weirich19}, repel~\cite{Moerman17}, or avoid crossing each other's trails~\cite{Kruger16}.

Robust active behavior and potential biocompatibility~\cite{Izri14} of active droplets makes them compelling building blocks for active matter engineering. Accordingly, substantial theoretical effort is aimed at developing reliable models of active droplets~\cite{Herminghaus14, Izri14, Yoshinaga17, Zwicker17, Morozov19a, Morozov19b, Morozov19c, Lippera20}. State-of-the-art models attribute spontaneous motion of active droplets to the Marangoni effect, that is, an interfacial flow emerging due to a chemically-induced gradient of the droplet surface tension~\cite{Herminghaus14, Izri14, Yoshinaga17, Zwicker17, Morozov19a, Lippera20}. For simplicity, diffusion-controlled surfactant sorption kinetics is typically postulated, when surfactant concentration at the interface is proportional to the bulk concentration~\cite{Izri14, Yoshinaga17, Zwicker17, Morozov19a, Morozov19b, Morozov19c}. Existing models also assume that the chemical reaction at the interface is a 0th~\cite{Izri14, Morozov19a, Morozov19b, Morozov19c, Lippera20} or 1st~\cite{Yoshinaga17, Zwicker17} order reaction.

We argue that existing models impose a set of restrictive assumptions that are violated when surfactant concentration exceeds the critical micelle concentration (CMC). At the same time, most observations of droplet motility due to micellar solubilization were conducted at surfactant concentrations far exceeding the CMC~\cite{Izri14, Kruger16, Moerman17, Suga18}. In this case, concentration of the monomers available for adsorption at the droplet interface should always be close to the CMC and it is unclear how an active droplet establishes the concentration gradient necessary for the onset of the Marangoni flow. Moreover, high values of the surfactant concentration required for the onset of the droplet motion hint that kinetics of the corresponding reaction is highly nonlinear, as it is supposed to be in the case of micelle assembly~\cite{Karapetsas11}.

In this paper, we construct a theoretical model linking the spontaneous motion of a slowly dissolving microdrop with the specifics of surfactant sorption and micelles production at its interface. Our goal is to specifically consider micellar solubilization in a highly concentrated surfactant solution and identify: (a) how Marangoni flows may emerge at high surfactant concentrations, (b) what is the effect of nonlinear micelle assembly kinetics on the droplet dynamics, and (c) what is the effect of surfactant aggregates on the behavior of active droplets. The paper is organized as follows. We formulate the model and outline its basic features in Sec.~\ref{problem}. In Sec.~\ref{numerical}, we present the results of our numerical simulations, while details of our numerical methods and supporting asymptotic analysis are described in Appendices~\ref{asymptotic} and~\ref{method}. Finally, we discuss our findings in Sec.~\ref{discussion}.

%%%%%%%%%%%%%%%%%%%%
% Problem statement
%%%%%%%%%%%%%%%%%%%%
\section{Model formulation}
\label{problem}
\begin{figure}
  \centering
  \includegraphics[scale=0.4]{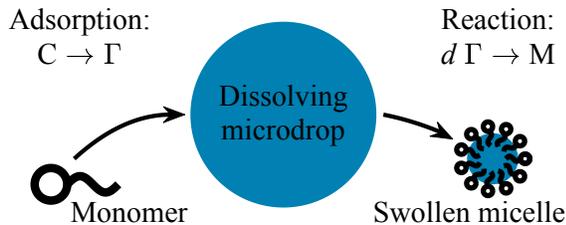}
  \caption{
    A spherical microdroplet gradually dissolving in the bulk of a surfactant solution.
    In the course of the dissolution, surfactant monomers adsorbed at the droplet interface
    form swollen micelles that carry away a small portion of liquid from the drop.
  }
  \label{fig1}
\end{figure}

%\textbf{Model formulation}
We consider a spherical droplet that is submerged in the bulk of a surfactant solution and undergoes gradual micellar solubilization, as sketched in Fig.~\ref{fig1}. When surfactant concentration exceeds the CMC, there should be at least three distinct chemical species involved in the solubilization process. The first are surfactant monomers; they are adsorbed at the droplet interface and serve as building blocks for the second species, swollen micelles, that carry a minuscule amount of fluid from the drop and, thus, constitute the product of the dissolution process. In addition, excess of monomers in the bulk leads to formation of regular micelles that are the third species contributing to the phenomena we aim to model. When droplet dissolution is slow, it is natural to assume that monomers and regular micelles remain in equilibrium, i.e., regular micelles are constantly assembled or disassembled to keep the bulk concentration of monomers constant and equal to the critical micelle concentration, $C_{CMC}$. This assumption allows us to disregard the bulk transport of monomers and regular micelles.
%in such a way that bulk concentration of monomers remains constant,
%\begin{equation}
%  \label{prob_c}
%  C^* = C_{CMC},
%\end{equation}
%where $C_{CMC}$ denotes critical micelle concentration and $*$ is used to mark dimensional variables. Note that assumption~\eqref{prob_c} effectively allows us to disregard the bulk transport of monomers and regular micelles.

In contrast to the regular micelles, swollen micelles are filled with liquid and can not be easily disassembled, as their dissociation requires creation of a tiny droplet with a ``clean'' interface (which is equivalent to reversing the micellar solubilization altogether). Based on this, we postulate that swollen micelles can not be disassembled and forever dwell in the bulk fluid, where their transport is governed by the advection-diffusion equation,
\begin{equation}
  \label{prob_m_trans}
  \partial_{t^*} M^* + \textbf{u}_o^* \cdot \nabla M^* = D_M \nabla^2 M^*,
\end{equation}
where $*$ is used to mark dimensional variables, $t^*$ is time, $M^*$ denotes the concentration of swollen micelles, $\textbf{u}_o^*$ is the flow velocity outside of the drop, and $D_M$ is the swollen micelles diffusivity. Since typical size of an active microdrop in experiments is $\sim 10 \, \mu$m~\cite{Izri14, Kruger16, Moerman17, Suga18}, we may safely neglect inertia and employ Stokes equations to describe the flow field within, $\textbf{u}_i^*$, and around the drop, $\textbf{u}_o^*$. We also emphasize that in our model neither of the chemical species may penetrate into the droplet.

Following Karapetsas~\textit{et al.}, we model micelles production rate at the droplet interface as~\cite{Karapetsas11},
\begin{equation}
  \label{prob_m_react}
  j_r^* = K_r \left( \Gamma^* \right)^d,
\end{equation}
where $K_r$ is the rate constant, $\Gamma^*$ denotes concentration of adsorbed monomers, and $d$ is an integer corresponding to the preferred number of monomers in a swollen micelle~\cite{Hunter91}. We assume that swollen micelles are not accumulated at the interface.
%, i.e., their desorption is sufficiently fast. 
In this case, desorption flux of swollen micelles at the droplet interface yields,
\begin{equation}
  \label{prob_m_prod}
  -D_M \partial_{r^*} M^* = j_r^*
  \qquad \text{at  } r^* = R,
\end{equation}
where $R$ is the droplet radius that is assumed to be constant, as the time of total droplet dissolution is exceedingly large compared to the typical timescale of the fluid flow.

Production of swollen micelles at the droplet interface is sustained by adsoprtion of surfactant monomers from the bulk. Here we employ a novel model of sorption kinetics that takes swollen micelles into account. Specifically, experiments reveal that self-propelling active droplets avoid crossing each other's paths~\cite{Kruger16}. It is hypothesized that this feature is due to the trail of swollen micelles that is left behind a self-propelling active drop. We assume that swollen micelles may interfere with the distribution of the monomers near the drop and, to incorporate this effect into the model, postulate that the sorption flux, $j_s$, depends on micelles concentration as follows,
\begin{equation}
  \label{prob_ad_flux}
  j_s^* = K_a e^{-\blu{\alpha_m} M^*} C_{CMC}
  \qquad \text{at  } r^* = R,
\end{equation}
where $K_a$ and \blu{$\alpha_m$} are adsorption and inhibition coefficients, respectively. In essence, Eq.~\eqref{prob_ad_flux} posits that micelles act as an adsorption inhibitor. \blu{This assumption is based on an intuition that diffusive transport of swollen micelles (i.e., large aggregates) away from the droplet interface is not efficient, while advection mainly redistributes swollen micelles along the interface. For instance, in the case of a self-propelling droplet, all of the produced swollen micelles will be pushed towards the back pole of the moving drop, creating a swollen micelle-rich zone. Since, swollen micelles can not be easily disassembled, they simply occupy space in the sublayer adjacent to the droplet interface and, thus, may partially block the adsorption of monomers.} We further assume that regular micelles do not inhibit adsorption, since regular micelles may be disassembled to compensate for the shortage of monomers. We also note that desorption is disregarded in Eq.~\eqref{prob_ad_flux} for simplicity.
% We argue that this mechanism may be especially relevant in solutions of ionic surfactants, since in this case both micelles and the drop acquire electrical charge. Charged micelles must be encapsulated in an electrical double layer that effectively increases the volume occupied by each micelle.

Monomers adsorbed at the droplet interface are continually consumed to produce swollen micelles. In this situation, although the CMC is reached in the bulk, amount of adsorbed monomers may still be insufficient for the formation of a continuous monolayer that hinders interfacial mobility~\cite{Cui13}. Therefore, we assume that monomer mobility along the interface is unrestricted and model their interfacial transport using a 2D advection-diffusion equation,
\begin{equation}
  \label{prob_gamma_trans}
  \partial_{t^*} \Gamma^* + \nabla_2 \cdot \left( \textbf{u}_2^* \Gamma^* \right) 
    = D_\Gamma \nabla_2^2 \Gamma^* + j_s^* - d \, j_r^*,
\end{equation}
where $\nabla_2$ and $\textbf{u}_2^*$ denote the interfacial gradient operator and interfacial flow velocity, respectively, while $D_\Gamma$ is the interfacial diffusivity. Accordingly, we also assume that flow velocity is continuous at the droplet interface.

Since we postulate that interfacial mobility is unrestricted and neglect possible adsorption of micelles, droplet surface tension may only depend on the interfacial concentration of monomers,
\begin{equation}
  \label{prob_marangoni}
  \gamma = \gamma_0 - \gamma_\Gamma \left( \Gamma^* - \Gamma_0 \right),
\end{equation}
where $\gamma_0$, $\Gamma_0$, and $\gamma_\Gamma$ are constants. Uneven surface tension contributes to the balance of stresses at the droplet interface via the Marangoni effect. Note that the droplet is assumed spherical at all times and the balance of normal stresses may be disregarded. 
%The balance of tangential stresses reads,
%\begin{equation}
%  \label{prob_stress}
%  \textbf{n} \cdot \left( \boldsymbol \tau_i^* 
%    - \boldsymbol \tau_o^* \right) \cdot \textbf{t}
%      = \textbf{t} \cdot \nabla \gamma
%  \qquad \text{at  } r^* = R,
%\end{equation}
%where $\boldsymbol \tau^*$ denotes viscous stress tensor.

Finally, when the coordinate system is co-moving with the drop, far away from the drop the flow is unidirectional, $\textbf{u}_o^* = -\textbf{U}_\infty^\blu{*}$, and swollen micelles concentration vanishes, $M^* = 0$. Here, $\textbf{U}_\infty^\blu{*}$ denotes droplet self-propulsion velocity that is determined from the condition that the total force acting on the drop is equal to $\textbf{0}$.

To obtain dimensionless form of the model, we use $R$ and $\Gamma_{CMC} \equiv K_a C_{CMC} R^2 / D_{\Gamma}$ as a unit of distance and interfacial concentration, respectively. Then we introduce a unit velocity, ${V \equiv \gamma_\Gamma \Gamma_{CMC} / \eta_o}$, a unit of time, ${R / V}$, and a unit of the swollen micelles concentration, ${K_r R \Gamma_{CMC}^d / D_M}$. Resulting dimensionless model equations are shown in Appendix~\ref{dimensionless_prob} and include six parameters: swollen micelle size, P{\'e}clet number, diffusivity contrast, viscosity contrast, and dimensionless inhibition and rate constants, respectively,
\begin{align}
  \label{prob_params}
  & d, \qquad \pe \equiv \frac{V R}{D_M}, \qquad 
  D \equiv \frac{D_M}{D_\Gamma}, \qquad
  \eta \equiv \frac{\eta_i}{\eta_o}, \\
  & k_m \equiv \blu{\frac{\alpha_m K_r R \Gamma_{CMC}^d}{D_M}}, \quad
  k_r \equiv \frac{K_r R^2 \Gamma_{CMC}^{d-1}}{D_\Gamma}.
\end{align}
Note that in experiments, P{\'e}clet number is typically estimated based on the droplet self-propulsion velocity, $U_\infty\blu{^* \equiv V U_\infty}$. In our model, this ``experimental'' P{\'e}clet number can be obtained as, $\pe_\text{exp} \equiv \pe \, U_\infty$.

%%%%%%%%%%%%%%%%%%%%%%%%
% Numerical simulations
%%%%%%%%%%%%%%%%%%%%%%%%
\section{Numerical simulations}
\label{numerical}
For simplicity, we simulate the droplet dynamics in the case of axisymmetric flow and concentration fields. To solve the model equations
%~\eqref{prob_m_trans}-\eqref{prob_stokes},~\eqref{prob_m_prod},~\eqref{prob_gamma_trans},~\eqref{prob_stress}-\eqref{prob_u_cont} 
numerically,
%Appendix~\ref{method}. 
we follow Ref.~\citenum{Morozov19b} and employ truncated expansions in Legendre harmonics.
% to approximate the flow and concentration fields around the droplet. 
Our simulations involve two steps: (a) the time-marching procedure is used to reach a steady flow regime around the droplet, (b) the result of time-marching then serves as an initial condition for the method of natural continuation that we use to obtain steady flow and concentration fields for different values of the problem parameters. We then numerically assess the linear stability of these steady flows. 
%For each set of parameter values, numerical solution is refined by means of Newton's iterations. We then employ numerical estimation of the Jacobian of the model equations to assess linear stability of the obtained steady flows. 
Detailed description of our numerical procedure in provided in Appendix~\ref{method}.

\subsection{Spontaneous droplet self-propulsion requires small diffusivity contrast}
Similarly to the models from Refs.~\citenum{Rednikov94, Michelin13b, Morozov19a, Morozov19b}, our model features a symmetry-breaking instability that results in the spontaneous self-propulsion of the droplet. This instability is due to a positive feedback mechanism enabled by advection of reagents in the bulk fluid. Specifically, advection carries swollen micelles away from the front of the moving droplet, thus enhancing the adsorption of surfactant monomers (as per Eq.~\eqref{prob_ad_flux}); in turn, increased surfactant adsorption causes a decrease in interfacial tension at the droplet front, resulting in the Marangoni flow that further boosts advection.

Unlike the models from Refs.~\citenum{Rednikov94, Michelin13b, Morozov19a, Morozov19b}, our model also accounts for the advection of adsorbed surfactant monomers along the droplet interface, Eq.~\eqref{prob_gamma_trans}. Physical intuition suggests that interfacial advection should homogenize the interfacial concentration of the surfactant, dampen the Marangoni flow, and, therefore, hinder the droplet self-propulsion. Consequently, a droplet may self-propel only when the bulk advection ``outperforms'' its interfacial counterpart. \blu{Since the velocity scale of the bulk and interfacial flow is the same, the difference in efficiency of advective transport must emerge due to different diffusivity of swollen micelles and adsorbed monomers which is quantified by the diffusivity contrast $D$.}
%We quantify the relative efficiency of the bulk and interfacial transport by the diffusivity contrast, $D$. 
In Appendix~\ref{asymptotic}, we use linear stability analysis to obtain the maximal value of $D$ that allows for the droplet self-propulsion. Specifically, we demonstrate that the critical P{\'e}clet number for the onset of self-propulsion is given by,
\begin{equation}
  \label{asymp_pe1_main}
  \pe_1 = \dfrac{2 (2 + 3 \eta) \left( 4 + k_r d^2 \kappa^{1-1/d} \left( 2 + k_m \kappa \right) \right)}
      {k_m k_r d \, \kappa^2 - 8 D \kappa^{1/d}},
\end{equation}
where ${\kappa \equiv W \left( k_m / ( d \, k_r ) \right) / k_m}$ and $W(x)$ is Lambert $W$ function. Since the onset of self-propulsion requires a finite and positive $\pe_1$, equation~\eqref{asymp_pe1_main} implies that self-propulsion may only emerge when ${D < k_m k_r d \, \kappa^{2-1/d} / 8}$. To comply with the latter requirement, we assume ${D = 10^{-3}}$ in our numerical simulations. We argue that a value of $D \ll 1$ is realistic, as it implicates that the interfacial diffusion of individual surfactant molecules is faster than the bulk diffusion of large aggregates.
% (i.e., swollen micelles).
\blu{Indeed, Ref.~\citenum{Valkovska00} estimates surface diffusivity of dodecanol molecules at the air-nitrobenzene interface as ${D_{\Gamma} \sim 10^{-9}}$~m$^2$/s. On the other hand, diffusivity of CO$_2$-swollen micelles in a polyol solution was measured to be ${D_M \sim 10^{-12}}$~m$^2$/s, or roughly $10^3$ times lower than the diffusivity of individual CO$_2$ molecules~\cite{Knoll19}.}

\subsection{Two types of active behavior: steady self-propulsion and pumping}
\begin{figure*}
  \centering
  \includegraphics[scale=0.52]{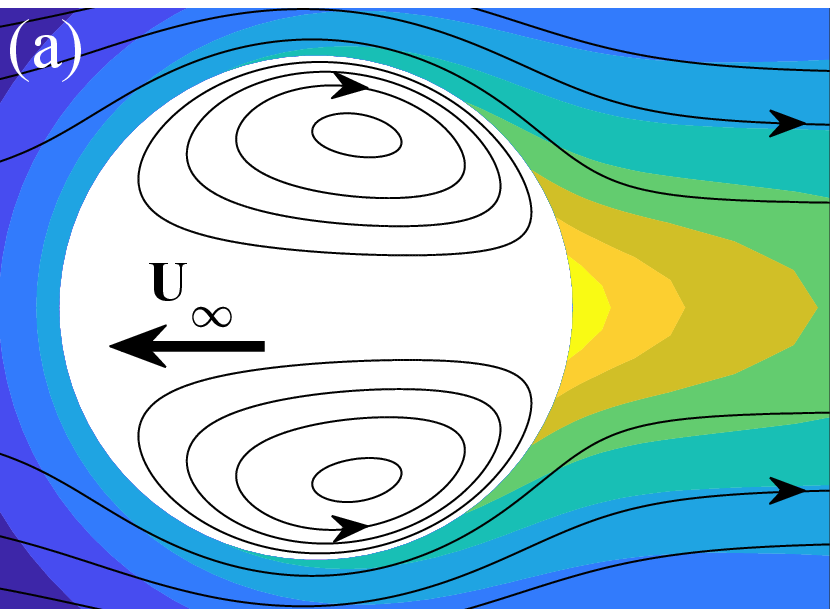}
  \qquad \;
  \includegraphics[scale=0.52]{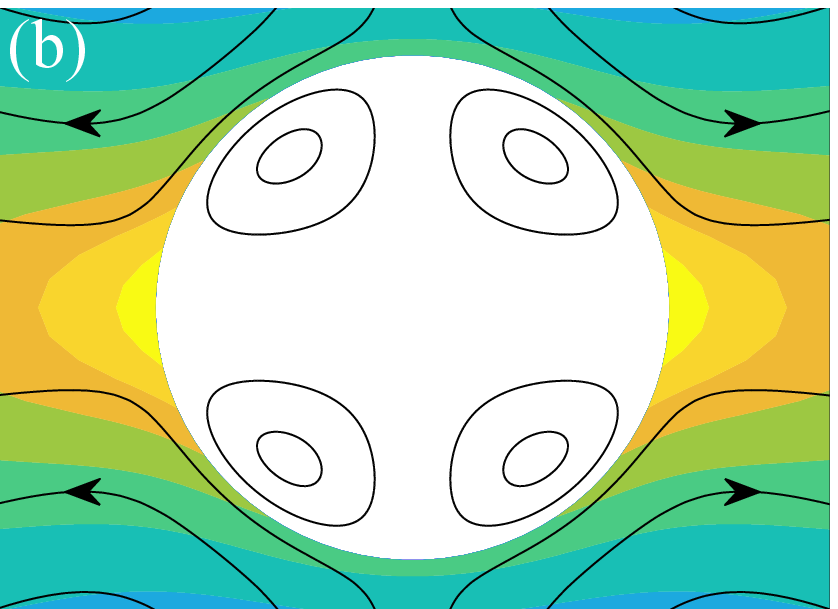}
  \; \; \;
  \includegraphics[scale=0.52]{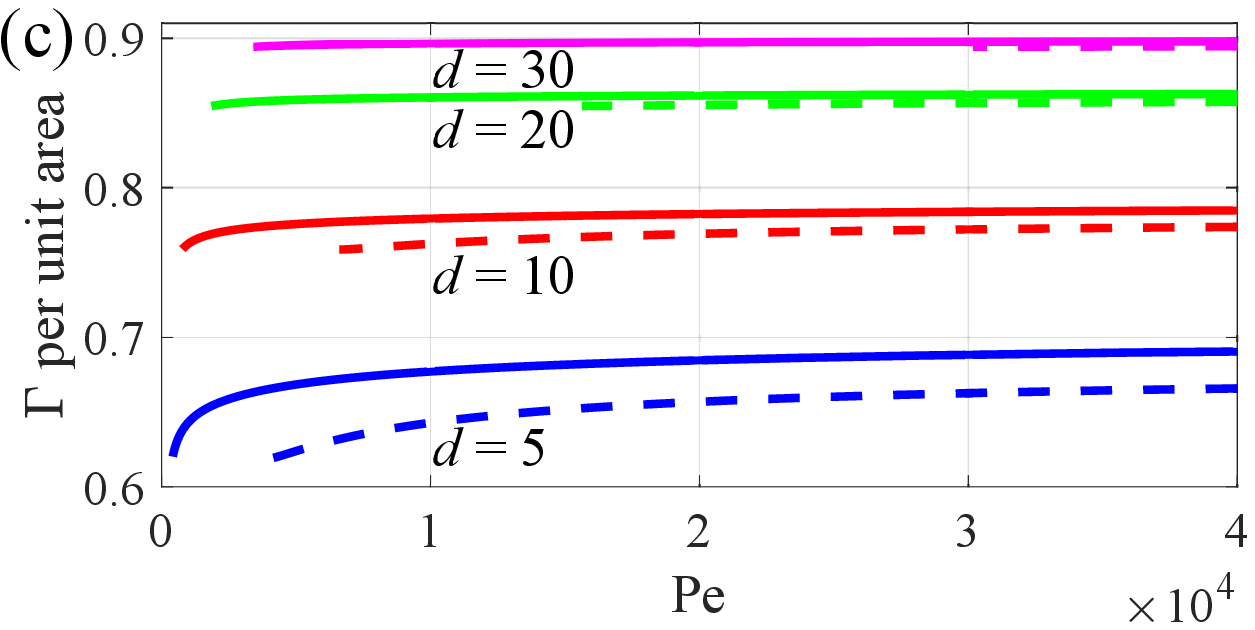}
  \caption{
    Two types of steady behavior of a dissolving microdrop:
    (a)~droplet self-propelling with constant velocity $\textbf{U}_\infty$ 
    and (b)~stationary droplet pumping fluid around in a symmetric manner.
    (c) Average concentration of adsorbed surfactant, $\Gamma$, per unit area of the droplet
    for $\eta = 1$, $D = 10^{-3}$, $k_m = 10$, $k_r = 1$, and various swollen micelle sizes $d$.
    Solid lines correspond to steady self-propulsion;
    dashed lines represent pumping flows.
    %\blu{Note how $\Gamma$ remains constant in a wide range of $\pe$,
    %consistent with experimental observations in Ref.~\cite{Suga18}.}
  }
  \label{figFlow}
\end{figure*}

In our time-marching simulations, spontaneous onset of the flow around an active droplet resulted in one of the two steady regimes \blu{shown in Fig.~\ref{figFlow}. Depending on the initial condition of the particular simulation,} the droplet was either self-propelling with a constant velocity or pumping the fluid around in a symmetric manner. Therefore, in what follows we focus on the properties of these two steady regimes.

We notice that in both self-propelling and pumping regimes, droplet dissolution rate per unit area remains almost constant in a wide range of P{\'e}clet numbers, $\pe$. Recall, that as per Eq.~\eqref{prob_m_prod}, dissolution rate is a function of the adsorbed surfactant concentration, $\Gamma$; and in Fig.~\ref{figFlow}c, we illustrate that $\Gamma$ per unit area remains fixed in a wide interval of $\pe$. This observation is consistent with experiments of Suga~\textit{et al.}, who observed linear decrease in diameter of dissolving active droplets~\cite{Suga18}. In turn, linear decrease in diameter implies quadratic decay in the reaction front area and, thus, roughly constant reaction rate per unit area.

\begin{figure*}
  \centering
  \includegraphics[scale=0.45]{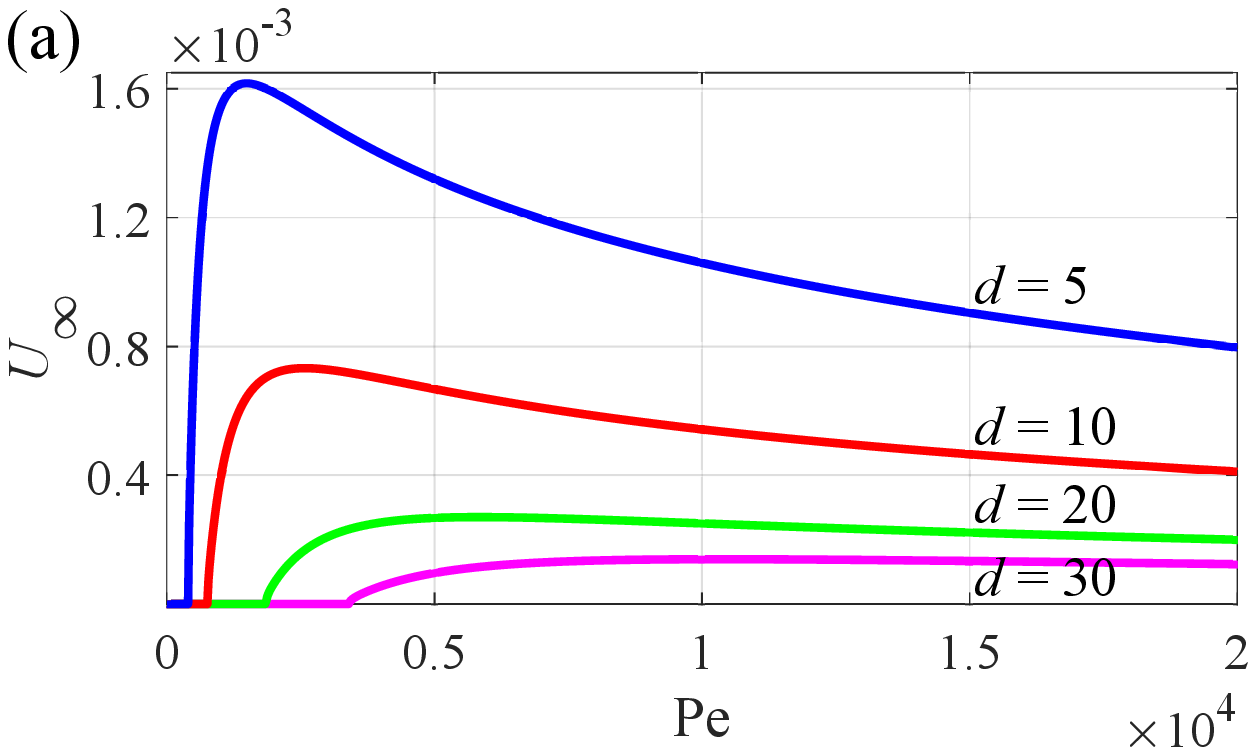}
  \quad
  \includegraphics[scale=0.45]{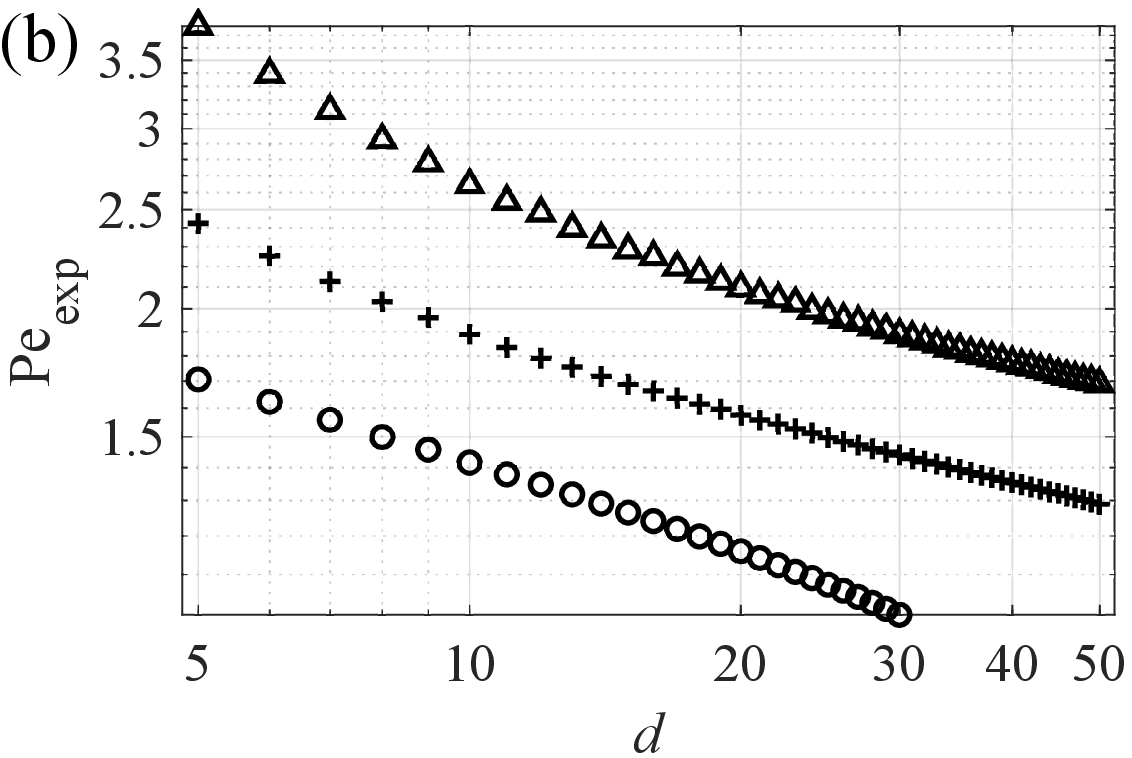}
  \quad
  \includegraphics[scale=0.45]{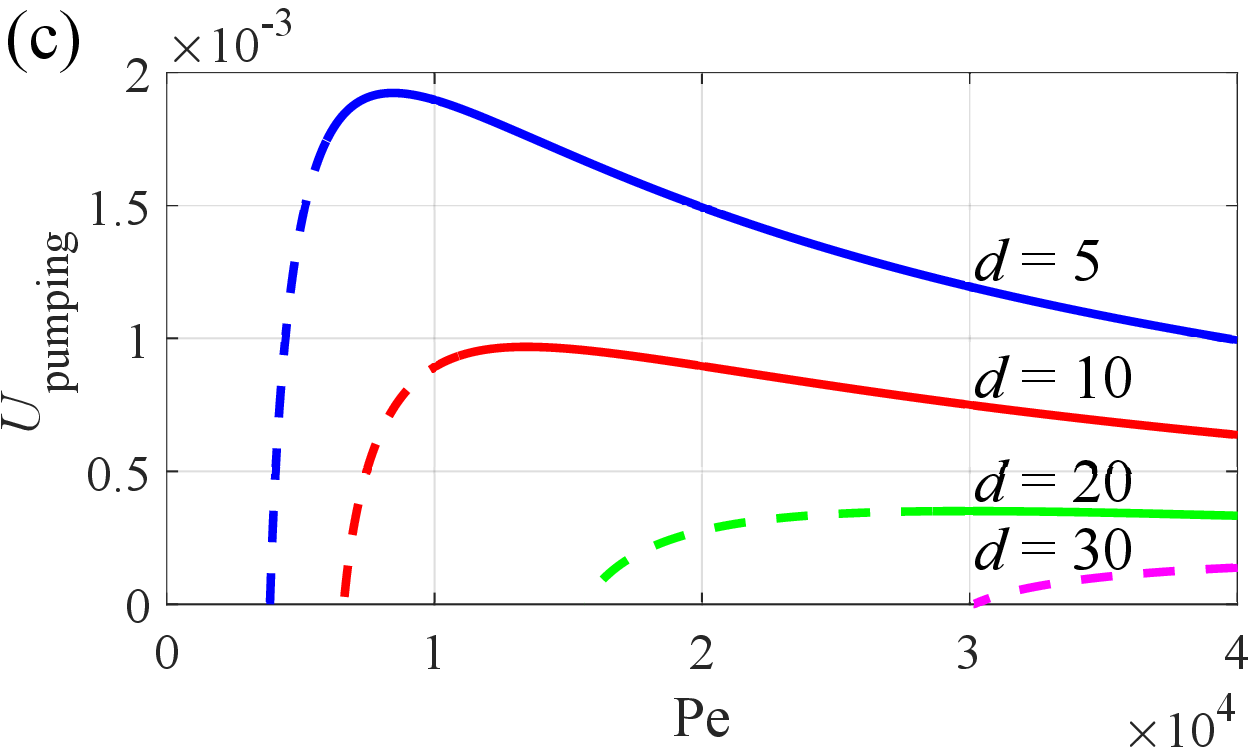}
  \caption{
    (a) Steady self-propulsion velocity versus P{\'e}clet number
    for $\eta = 1$, $D = 10^{-3}$, $k_m = 10$, $k_r = 1$, and various swollen micelle sizes.
    %Peak of dimensionless self-propulsion velocity was also observed in Refs.~\cite{Michelin13b, Morozov19b}.
    (b) Drop velocity-based P{\'e}clet number, ${\pe_\text{exp} \equiv \pe \, U_\infty}$,
    corresponding to the peak of the dimensionless velocity in (a):
    ``+'' -- $k_m = 10$ and $k_r = 1$;
    ``$\triangle$'' -- $k_m = 50$ and $k_r = 1$;
    ``$\circ$'' -- $k_m = 10$ and $k_r = 5$.
    Other parameters: $\eta = 1$ and $D = 10^{-3}$.
    Peak velocity for $k_m = 10$, $k_r = 5$, and $d > 30$ lies beyond 
    $\pe = 40000$ that was the maximal value of the P{\'e}clet number considered in this paper.
    (c) Steady symmetric pumping flow velocity
    for $\eta = 1$, $D = 10^{-3}$, $k_m = 10$, $k_r = 1$, and various swollen micelle sizes.
    Dashed lines correspond to unstable pumping flows;
    solid lines represent stable pumping flows.
    \blu{Pumping flow velocity is defined as the maximal velocity of the second squirming mode in Eq.~\eqref{prob_psi_o}.}
  }
  \label{figDropU}
\end{figure*}

Similarly to the models in Refs.~\citenum{Michelin13b, Morozov19b}, droplet self-propulsion velocity reaches a peak at a certain value of the P{\'e}clet number, as illustrated in Fig.~\ref{figDropU}a. This peak corresponds to an optimal configuration of the advective flow carrying swollen micelles from the front towards the back of the propelling drop. Since dimensionless velocity scales with the droplet size in our model, it is hard to argue that a similar peak should be observed in experiments. However, as the peak is positioned far from the instability threshold, it is clear that in an experimental setting this point should correspond to a well-develped self-propelling flow. Since in experiments the P{\'e}clet number is typically estimated based on the droplet self-propulsion velocity, we utilize the peak velocity to assess the typical ``experimental'' P{\'e}clet number, $\pe_\text{exp} \equiv \pe \, U_\infty$, sufficient for the droplet self-propulsion. As we illustrate in Fig.~\ref{figDropU}b, our simulations yield a typical value of \blu{$\pe_\text{exp} \sim 1$}, regardless of the values of reaction coefficents $k_m$, $k_r$, and swollen micelle size,~$d$.

\subsection{Self-propelling and pumping states may be multistable}
\begin{figure*}
  \centering
  \includegraphics[scale=0.47]{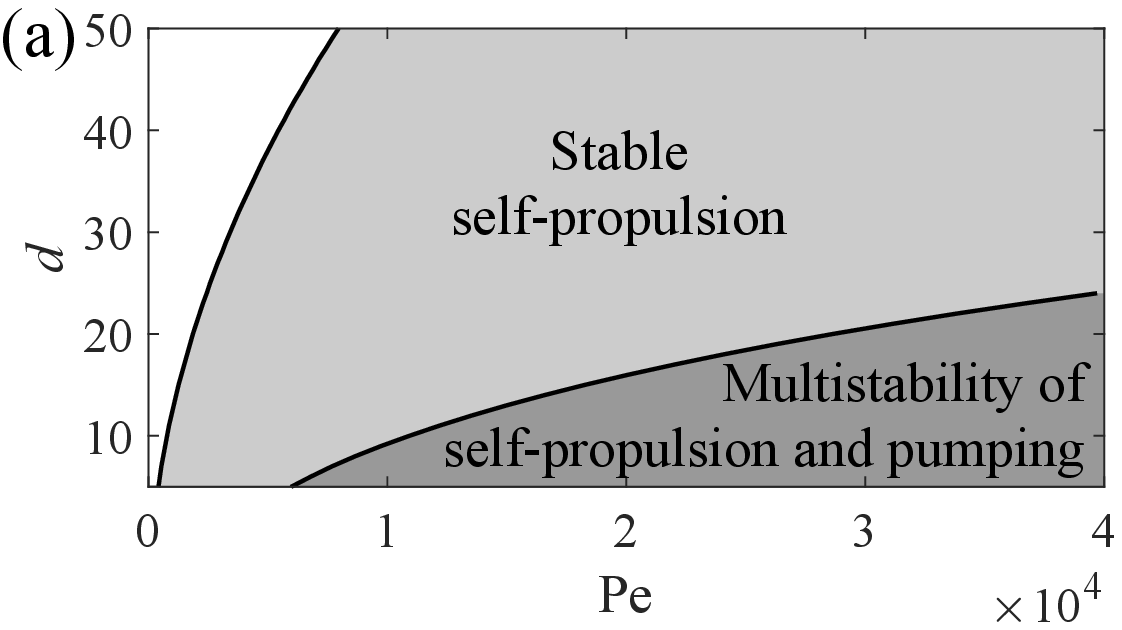}
  \qquad
  \includegraphics[scale=0.47]{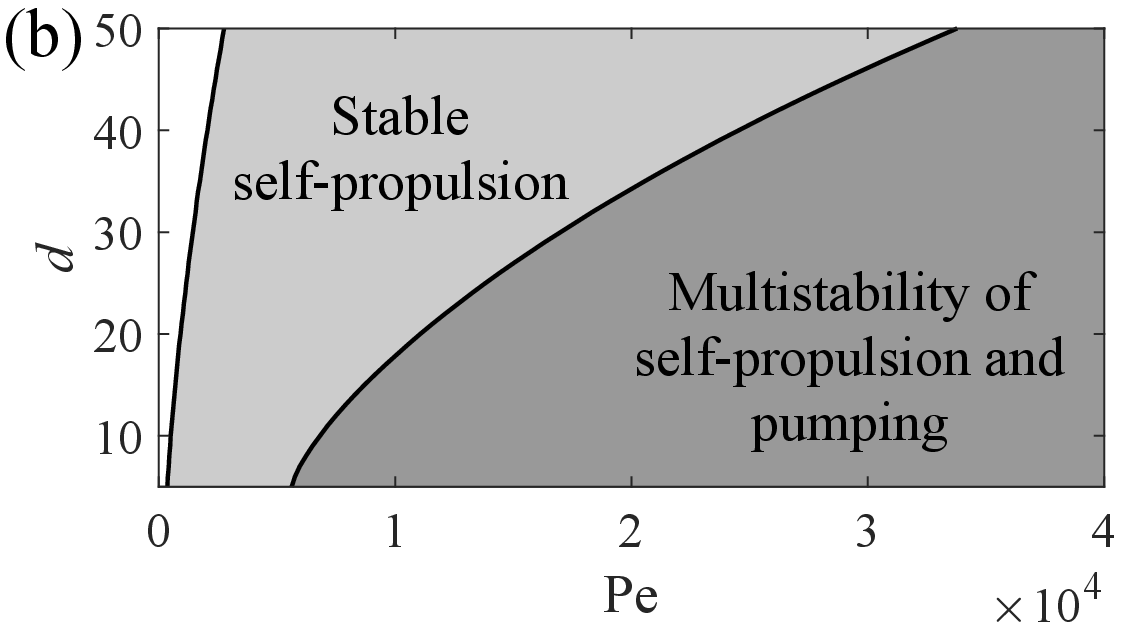}
  \qquad
  \includegraphics[scale=0.47]{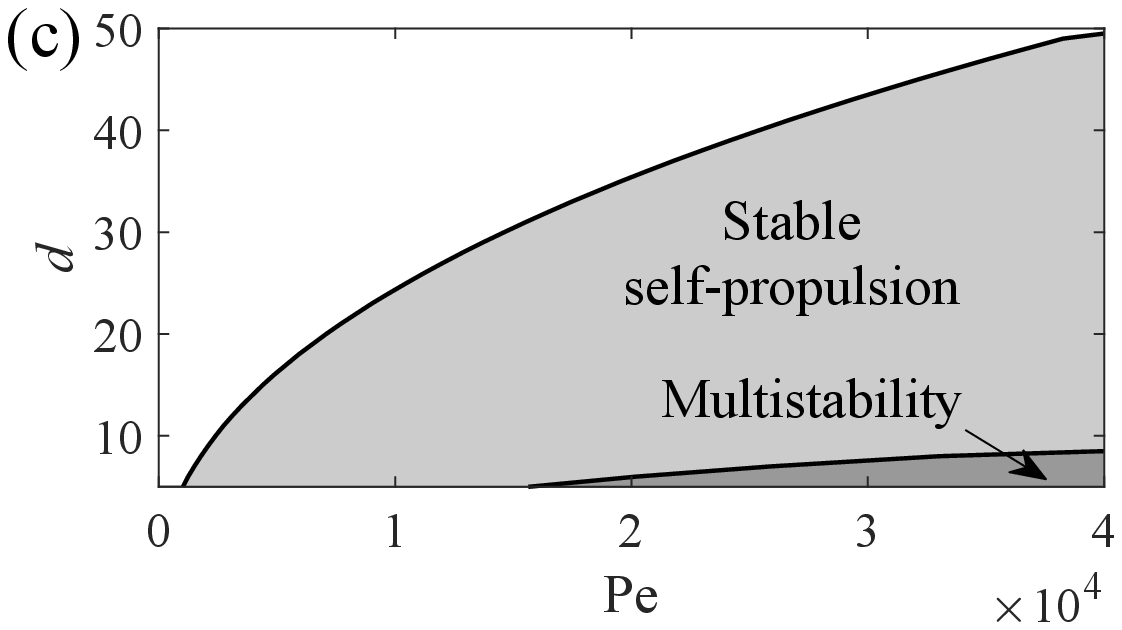}
  \caption{
    Stability maps of the steady flow regimes around an active microdroplet.
    \blu{Self-propulsion with constant velocity (Fig.~\ref{figFlow}a) is stable in the light gray area;
    both self-propulsion (Fig.~\ref{figFlow}a) and symmetric pumping (Fig.~\ref{figFlow}b)
    are stable in the dark gray area;
    no spontaneous onset of the flow was observed in the white area.}
    Panel (a): $k_m = 10$ and $k_r = 1$;
    (b): $k_m = 50$ and $k_r = 1$;
    (c): $k_m = 10$ and $k_r = 5$.
    Other parameters: $\eta = 1$ and $D = 10^{-3}$.
  }
  \label{figMap}
\end{figure*}

Our simulations indicate that both self-propelling and pumping flow regimes may be \blu{linearly} stable for a given set of the problem parameters. In particular, we observe that pumping states bifurcate subcritically, but may regain stability at a higher value of the P{\'e}clet number, as illustrated in Fig.~\ref{figDropU}c. In Fig.~\ref{figMap}, we further demonstrate that the multistability of self-propelling and pumping states exists in a wide range of problem parameters. Our results suggest that multistability is promoted by the adsorption inhibition, as higher value of the inhibition coefficient, $k_m$, results in wider region of multistability. In turn, higher reaction rate, $k_r$, destabilizes pumping states, suggesting that intense reaction makes it harder to establish a symmetric surfactant distribution required for a stable pumping flow.

%%%%%%%%%%%%%
% Discussion
%%%%%%%%%%%%%
\section{Discussion}
\label{discussion}
In this paper, we propose a novel model of an active microdroplet that undergoes gradual micellar solubilization in the bulk of a concentrated surfactant solution. To the best of our knowledge, this model is the first attempt to explain how the Marangoni effect may emerge at surfactant concentrations exceeding the CMC, and what is the role of surfactant aggregates in the dynamics of active microdrops. Our model is based on two key assumptions: (a) swollen micelles released during the dissolution occupy space in the sublayer adjacent to the droplet interface and, thus, may inhibit surfactant adsorption, Eq.~\eqref{prob_ad_flux}; (b) solubilization is fast enough to prevent the formation of a monolayer of surfactant monomers at the interface and interfacial transport may be modeled by advection-diffusion equation, Eq.~\eqref{prob_gamma_trans}.

Our numerical simulations suggest that regardless of the values of adsorption and reaction coefficients, droplet self-propulsion in experiments should be observed for \blu{${\pe_\text{exp} \sim 1}$}, as illustrated in Fig.~\ref{figDropU}b. Modest estimated values of $\pe_\text{exp}$ are in stark contrast with the values of the P{\'e}clet number, $\pe$, used in Figs.~\ref{figFlow}c, \ref{figDropU}a, \ref{figDropU}c, and~\ref{figMap}. This contrast is due to fundamentally different physical meaning of $\pe$ and $\pe_\text{exp}$: $\pe_\text{exp}$ is based on the droplet velocity and quantifies how efficiently the flow stirs micelles in the bulk fluid \blu{(and the estimation ${\pe_\text{exp} \sim 1}$ means that advective and diffusive transport are about equally important)}, while $\pe$
%, defined in Eq.~\eqref{prob_params}, 
involves intensity of the droplet surfactant intake and, thus, is a measure of the droplet chemical activity. 
%We argue, that it is not a coincidence, that a number of models of active droplet utilize the P{\'e}clet number based on the doroplet activity, as the bifurcation parameter~Refs.~\cite{Michelin13b, Izri14, Yoshinaga17, Morozov19a, Morozov19b, Morozov19c, Lippera20}.

Our model also highlights the antagonism between the bulk and interfacial advection. Bulk advection carries swollen micelles away from the front of a moving droplet, thus enhancing surfactant adsorption. In turn, increased adsorption results in the gradient of adsorbed monomers that generates the Marangoni flow and promotes the instability. On the other hand, interfacial advection homogenizes the interfacial surfactant concentration, hinders the Marangoni flow, and suppresses the instability. Consequently, droplet chemical activity may result in the flow only when bulk advection ``outperforms'' its interfacial counterpart. In Appendix~\ref{asymptotic}, we demonstrate that the latter is possible when interfacial diffusion of monomers is more efficient than the bulk diffusion of swollen micelles.

Finally, our key finding is that in the case of an axisymmetric flow and concentration fields, two qualitatively different types of droplet behavior may be stable for the same values of the problem parameters: steady self-propulsion (Fig.~\ref{figFlow}a) and steady symmetric pumping (Fig.~\ref{figFlow}b). \blu{In this situation, temporal evolution of the system may result in either of these two steady regimes, depending on the initial conditions of the model equations.} Although stability of the steady regimes shown in Fig.~\ref{figFlow} is not guaranteed in the absence of axial symmetry, we argue that they will retain their respective stable manifolds and, thus, appear at least as saddles in the phase space of a fully 3D problem. Simultaneous existence of two fixed points with stable manifolds is crucial, as the competition between these attractors may contribute to the large fluctuations of droplet self-propulsion velocity reported in Ref.~\citenum{Izri14}. We further argue that a vast majority of the contemporary models of active matter assume that individual active agents typically remain in vicinity of the self-propelling steady state~\cite{Marchetti13}. Our findings not only indicate that there is an alternative steady state, but also that the latter state possesses a stable manifold, so the active agents may be attracted to it. Therefore, we call for a creation of a novel class of active matter models that take possible bistability of the individual agents into account.

\section*{Conflicts of interest}
There are no conflicts to declare.

\section*{Acknowledgements}
This project has received funding from the European Union’s Horizon 2020 research and innovation programme under the Marie Sk{\l}odowska-Curie grant agreement No.~801505. M.M. is grateful to Prof. Fabian Brau and Prof. Laurence Rongy for the critical read of the manuscript, to Prof. S{\'e}bastien Michelin for enlightening discussions and rigorous attention to plots, and to the HPC team of ULB for the provided computational resources.

%%%%%%%%%%%%%
% Appendices
%%%%%%%%%%%%%
\appendix

\section{Dimensionless form of the model equations}
\label{dimensionless_prob}
Dimensional governing equations of the model read,
\begin{align}
  & \partial_{t^*} M^* + \textbf{u}_o^* \cdot \nabla M^* = D_M \nabla^2 M^*, \\
  \label{prob_stokes}
  & \nabla \cdot \textbf{u}_{i,o}^* = 0, \qquad
  \nabla P_{i,o}^* = \eta_{i,o} \nabla^2 \textbf{u}_{i,o}^*,
\end{align}
where $\textbf{u}^*$ is the flow velocity, $P^*$ is pressure, $\eta$ is the fluid viscosity, while subscripts $i$ and $o$ mark the corresponding quantity within and outside of the drop.

Dimensional boundary conditions at the droplet interface ($r^* = R$) read,
\begin{align}
  & \partial_{t^*} \Gamma^* + \nabla_2 \cdot \left( \textbf{u}_2^* \Gamma^* \right) 
    = D_\Gamma \nabla_2^2 \Gamma^* + j_s^* - d \, j_r^*, \\
  & \textbf{n} \cdot \left( \boldsymbol \tau_i^* 
    - \boldsymbol \tau_o^* \right) \cdot \textbf{t}
      = \textbf{t} \cdot \nabla \gamma, \\
  & \textbf{u}_i^* \cdot \textbf{t} = \textbf{u}_o^* \cdot \textbf{t}, \quad
    \textbf{u}_i^* \cdot \textbf{n} = \textbf{u}_o^* \cdot \textbf{n} = 0,
\end{align}
where $\nabla_2$ and $\textbf{u}_2^*$ denote the interfacial gradient operator and interfacial flow velocity, respectively, while $D_\Gamma$ is interfacial diffusivity, and $\boldsymbol \tau^*$ denotes viscous stress tensor.

We employ a coordinate system co-moving with the drop. In this case, dimensional boundary conditions far away form the drop read,
%We employ a coordinate system co-moving with the drop. In this case, far away from the drop the flow is unidirectional flow, while pressure and swollen micelles concentration is constant,
\begin{equation}
  \label{prob_far}
  \textbf{u}_o^* = -\textbf{U}_\infty^\blu{*}, \quad
  P_o^* = P_\infty, \quad
  M^* = 0 \quad
  \text{as  } r^* \rightarrow \infty,
\end{equation}
where $\textbf{U}_\infty^\blu{*}$ denotes droplet self-propulsion velocity that is determined from the condition that the total force acting on the drop is equal to $\textbf{0}$.

To obtain dimensionless form of the model, we use $R$ as a unit of distance and define a unit of time, ${R / V}$, based on the Marangoni flow velocity, $V \equiv \gamma_\Gamma \Gamma_{CMC} / \eta_o$, with $\Gamma_{CMC} \equiv K_a C_{CMC} R^2 / D_{\Gamma}$. Dimensionless pressure, viscous stresses, and concentrations $M$ and $\Gamma$, are defined, respectively, as,
\begin{align}
  \label{prob_dims}
  & P_i^* \rightarrow P_\infty + \frac{2 \gamma_0}{R} + \frac{\eta_i V}{R}, \qquad
  P_o^* \rightarrow P_\infty + \frac{\eta_o V}{R}, \\
  & \left( \boldsymbol \tau_i^*, \boldsymbol \tau_o^* \right)
    \rightarrow \frac{\eta_o V}{R}
      \left( \eta \boldsymbol \tau_i, \boldsymbol \tau_o \right), \\
  & M^* \rightarrow \frac{K_r R \Gamma_{CMC}^d}{D_M} M, \qquad
  \Gamma^* \rightarrow  \Gamma_{CMC} \Gamma,
\end{align}
where ${\eta \equiv \eta_i / \eta_o}$.

Dimensionless form of the governing equations
%~\eqref{prob_m_trans}-\eqref{prob_stokes} 
reads,
\begin{align}
  \label{prob_nondim_beg}
  & \nabla \cdot \textbf{u}_{i,o} = 0, \qquad
  \nabla P_{i,o} = \nabla^2 \textbf{u}_{i,o}, \\
  \label{prob_nondim_m_trans}
  & \pe \left( \partial_t M + \textbf{u}_o \cdot \nabla M \right) = \nabla^2 M.
\end{align}
Dimensionless boundary conditions at ${r = 1}$ may be written as,
\begin{align}
  & \textbf{u}_i \cdot \textbf{t} = \textbf{u}_o \cdot \textbf{t}, \qquad
  \textbf{u}_i \cdot \textbf{n} = \textbf{u}_o \cdot \textbf{n} = 0, \\
  & \partial_r M = -\Gamma^\blu{d}, \qquad
  \textbf{n} \cdot \left( \eta \boldsymbol \tau_i 
    - \boldsymbol \tau_o \right) \cdot \textbf{t}
      = -\textbf{t} \cdot \nabla \Gamma, \\
  \label{prob_nondim_end}
  & \pe D \big( \partial_t \Gamma + \nabla_2 \cdot \left( \textbf{u}_2 \Gamma \right) \big) 
    = \nabla_2^2 \Gamma + e^{-k_m M} - d \, k_r \Gamma^d.
\end{align}

\section{Axisymmetric Stokes flow past a spherical drop}
\label{stokes}
In this paper, we assume that the flow field is axisymmetric and introduce axisymmetric spherical coordinates ${\textbf{r} \equiv ( r, \theta )}$ with $r=0$ corresponding to the droplet center. In the case of an axisymmetric flow that is continuous at $r = 1$, Stokes equations~\eqref{prob_nondim_beg} admit a solution in the form of a superposition of orthogonal modes~\citep{Lamb45, Happel83, Leal07},
\begin{align}
  \label{prob_psi_i}
  & \Psi_i(t,\textbf{r}) = \frac{3}{4} a_1(t) \psi_{i,1}(\textbf{r}) 
    + \sum\limits_{n=2}^\infty a_n(t) \psi_{i,n}(\textbf{r}), \\
  \label{prob_psi_in}
  & \psi_{i,n}(\textbf{r}) = r^{n+1} \left( 1 - r^2 \right)
      \left( 1 - \mu^2 \right) L_n'(\mu), \\
  \label{prob_psi_o}
  & \Psi_o(t,\textbf{r}) = \sum\limits_{n=1}^\infty a_n(t) \psi_{o,n}(\textbf{r}), \\
  \label{prob_psi_on}
  & \psi_{o,n}(\textbf{r}) = \begin{cases}
    \left( 1 - r^3 \right) \left( 1 - \mu^2 \right) / \left( 2 r \right) \; & n = 1\\
    \left( 1 - r^2 \right) 
      \left( 1 - \mu^2 \right) L_n'(\mu) / r^n \; & n > 1
  \end{cases},
\end{align}
where $\Psi_i$ and $\Psi_o$ are the stream functions within and outside of the droplet, respectively, whereas $\psi_{i,n}$ and $\psi_{o,n}$ denote the $n$-th mode of the corresponding flow decomposition, $a_n(t)$ are time-dependent amplitudes, ${\mu \equiv \cos \theta}$, and $L_n$ denotes the $n$-th Legendre polynomial. It should be noted that orthogonal modes~\eqref{prob_psi_in} and~\eqref{prob_psi_on} with different $n$ correspond to qualitatively different flow regimes. For instance, $\psi_{o,1}$ represents the flow around a drop that self-propels with velocity ${\textbf{U}_\infty = a_1 \textbf{e}_z}$, where $\textbf{e}_z$ is a unit vector along the symmetry axis of the drop. In turn, $\psi_{o,2}$ describes a symmetric pumping flow regimes that does not feature droplet self-propulsion.

%%%%%%%%%%%%%%%%%%%%%%
% Asymptotic analysis
%%%%%%%%%%%%%%%%%%%%%%
\section{Asymptotic analysis}
\label{asymptotic}
It is easy to see that dimensionless problem formulated by Eqs.~\eqref{prob_nondim_beg}-\eqref{prob_nondim_end} admits the following trivial solution,
\begin{equation}
  \label{prob_base_state}
  \textbf{u}_i = \textbf{u}_o = \textbf{0}, \qquad
  M = \frac{\kappa}{r}, \qquad
  \Gamma = \kappa^{1/d},
\end{equation}
where $\kappa \equiv W \left( k_m / ( d \, k_r ) \right) / k_m$ and $W(x)$ is Lambert $W$ function.

We now employ matched asymptotic expansions to seek for the steady axisymmetric flow regimes that may emerge in vicinity of the motionless base state and, consequently, may be implemented as a time-independent asymptotic correction to the trivial solution~\eqref{prob_base_state}. Our approach closely follows the algorithm developed in Ref.~\citenum{Morozov19a}. Accordingly, here we keep technical discussion to minimum and focus on the physical implications of the results instead.

For a steady flow in vicinity of the base state~\eqref{prob_base_state}, amplitudes $a_n$ in Eqs.~\eqref{prob_psi_i} and~\eqref{prob_psi_o} should be time-independent and small. Accordingly, we expand $a_n$ in powers of $\epsilon$ as follows,
\begin{equation}
  \label{asymp_a}
  a_n = \epsilon a_n^{(1)} + \epsilon^2 a_n^{(2)} + \ldots,
\end{equation}
where $0 < \epsilon \ll 1$. \blu{Although velocity of the flow is small, it does not decay far away from the moving drop (recall that we use a coordinate system co-moving with the droplet). This non-vanishing flow makes it impossible to satisfy the far-field boundary conditions~\eqref{prob_far} in the framework of regular asymptotic expansion. Instead, matched asymptotic expansions are required in this case~\cite{Acrivos62}.}

The limit of weak advection around a spherical droplet allows for a composite steady solution of the advection-diffusion equation~\eqref{prob_nondim_m_trans} comprising a near field part, $N(\textbf{r})$, valid for ${r \ll 1/\epsilon}$, and a far field part, $F(\textbf{r})$, valid for ${r \gg 1}$~\citep{Acrivos62, Rednikov94, Morozov19a},
\begin{equation}
  M(\textbf{r}) = \begin{cases}
    N(\textbf{r}) \; & r \ll 1/\epsilon \\
    F(\textbf{r}) \; & r \gg 1
  \end{cases}.
\end{equation}
Following Ref.~\citenum{Morozov19a}, we expand $N(\textbf{r})$ and $F(\textbf{r})$ as follows,
\begin{align}
  \label{asymp_n_exp}
  & N(\textbf{r}) = \frac{\kappa}{r} + \epsilon N^{(1)}(\textbf{r}) 
    + \epsilon^2 N^{(2)}(\textbf{r}) + \ldots, \\
  \label{asymp_f_exp}
  & F(\boldsymbol \rho) = \epsilon F^{(1)}(\boldsymbol \rho)
    + \epsilon^2 F^{(2)}(\boldsymbol \rho) + \ldots,
\end{align}
where ${\boldsymbol \rho \equiv ( \rho, \theta ) = ( \epsilon r, \theta )}$ is the stretched radius vector. Finally, solution for the interfacial concentration $\Gamma$ must conform with expansion~\eqref{asymp_n_exp}, namely,
\begin{equation}
  \label{asymp_gamma_exp}
  \Gamma(\mu) = \kappa^{1/d} + \epsilon \Gamma^{(1)}(\mu) + \epsilon^2 \Gamma^{(2)}(\mu) + \ldots.
\end{equation}

\subsection{Linear analysis (terms at $\epsilon^1$)}
\label{asymp1}
We substitute expansions~\eqref{asymp_a} and~\eqref{asymp_n_exp}-\eqref{asymp_gamma_exp} into the dimensionless problem formulated by Eqs.~\eqref{prob_nondim_beg}-\eqref{prob_nondim_end} and collect the terms featuring equal powers of $\epsilon$. Naturally, collecting the terms at $\epsilon^1$ is equivalent to linearization of the dimensionless problem near the base state~\eqref{prob_base_state}. Since we only consider time-independent flows, the time-independent linear problem constituted by terms at $\epsilon^1$ corresponds to the situation when the small perturbations of the base state are neither growing nor decaying. By definition, this happens at the threshold of monotonic instability. Below, we identify this threshold and obtain the corresponding instability eigenmodes.

Linearity of the problem at $\epsilon^1$ allows us to treat each of the flow decomposition modes~\eqref{prob_psi_in},~\eqref{prob_psi_on} separately. To this end, we project the near field concentration of the swollen micelles and the concentration of adsorbed monomers onto the basis of Legendre polynomials,
\begin{align}
  \label{asymp_n1_exp}
  & N^{(1)}(\textbf{r}) = \sum\limits_{n=0}^\infty N_n^{(1)}(r) L_n(\mu), \\
  \label{asymp_gamma1_exp}
  & \Gamma^{(1)}(\mu) = \sum\limits_{n=0}^\infty \Gamma_n^{(1)} L_n(\mu).
\end{align}
Substitution of expansions~\eqref{asymp_n1_exp}-\eqref{asymp_gamma1_exp} in the problem at $\epsilon^1$ splits it into a set of problems for each of the Legendre modes. For instance, problem at $L_0$ reads,
\begin{align}
  \label{asymp_lin_p0_beg}
  & \left( \partial_{rr} + \frac{2}{r} \partial_r \right) N_0^{(1)} = 0, \\
  & \partial_r N_0^{(1)} = -d \, \kappa^{1-1/d} \Gamma_0^{(1)} \qquad \text{at  } r = 1, \\
  \label{asymp_lin_p0_end}
  & k_m e^{-k_m \kappa} N_0^{(1)} = -k_r d^2 \kappa^{1-1/d} \Gamma_0^{(1)} \qquad \text{at  } r = 1,
\end{align}
while problem at $L_1$ may be written as,
\begin{align}
  \label{asymp_lin_p1_beg}
  & \left( \partial_{rr} + \frac{2}{r} \partial_r - \frac{2}{r^2} \right) N_1^{(1)} 
    = \frac{\pe \, \kappa \, a_1^{(1)}}{r^2} \left( 1 - \frac{1}{r^3} \right), \\
  & \partial_r N_1^{(1)} = -d \, \kappa^{1-1/d} \Gamma_1^{(1)} \qquad \text{at  } r = 1, \\
  \label{asymp_lin_p1_mid}
  & 2 \Gamma_1^{(1)} = 3 ( 2 + 3 \eta ) a_1^{(1)} \qquad \text{at  } r = 1,
\end{align}
\begin{multline}
  \label{asymp_lin_p1_end}
  d \, k_r k_m \kappa N_1^{(1)} + \left( 2 + d \, \kappa^{1-1/d} \right) \Gamma_1^{(1)}
    \\ + 3 \pe D \kappa^{1/d} a_1^{(1)} = 0
      \qquad \text{at  } r = 1.
\end{multline}

Recall that $N^{(1)}$ given by superposition~\eqref{asymp_n1_exp} is only valid for ${r \ll 1/\epsilon}$. Far from the droplet, near field solution $N^{(1)}$ must match the far field solution $F^{(1)}$ that satisfies the leading order advection-diffusion equation in stretched coordinates,
\begin{equation}
  \label{asymp_f1_trans}
  \pe \, a_1^{(1)} \left( \mu, \, -\sqrt{1 - \mu^2} \right) \cdot \nabla_\rho F^{(1)} = \nabla_\rho^2 F^{(1)},
\end{equation}
where $\nabla_\rho$ denotes the gradient in terms of the stretched radius vector $\rho \equiv \epsilon r$.

Matching of the solution of Eq.~\eqref{asymp_f1_trans} with $N_0^{(1)}$ and $N_1^{(1)}$ obtained from Eqs.~\eqref{asymp_lin_p0_beg} and~\eqref{asymp_lin_p1_beg}, respectively, yields,
\begin{align}
  \label{asymp_n10_sol}
  & N_0^{(1)}(\textbf{r}) = b_0^{(1)} / r - \pe \, \kappa \, a_1^{(1)} / 2, \\
  \label{asymp_n11_sol}
  & N_1^{(1)}(\textbf{r}) = b_1^{(1)} / r^2 - \pe \, \kappa \, a_1^{(1)} ( 1 + 2 r^3 ) / ( 4 r^3 ), \\
  \label{asymp_f1_sol}
  & F^{(1)}(\boldsymbol \rho) = \kappa \, e^{-\pe a_1^{(1)} \rho \left( 1 + \mu \right) / 2} / \rho,
\end{align}
where constant coefficients $a_1^{(1)}$, $b_0^{(1)}$, and $b_1^{(1)}$ are to be determined from the boundary conditions. For simplicity, solution~\eqref{asymp_f1_sol} is written under assumption that ${a_1^{(1)} \geq 0}$ (i.e., droplet self-propels along $\textbf{e}_z$). This assumption may be made without loss of generality, since the problem under consideration is isotropic. Also note that matching condition for $N_n^{(1)}$ with $n > 1$ is simply ${\lim\limits_{r \rightarrow \infty} N_n^{(1)} = 0}$, as shown in Ref.~\citenum{Morozov19a}.

Substitution of the matched solutions~\eqref{asymp_n10_sol} and~\eqref{asymp_n11_sol} into boundary conditions~\eqref{asymp_lin_p0_end} and~\eqref{asymp_lin_p1_mid}-\eqref{asymp_lin_p1_end} yields a set of algebraic equations for $a_1^{(1)}$, $b_0^{(1)}$, $b_1^{(1)}$, $\Gamma_0^{(1)}$, and $\Gamma_1^{(1)}$. Solvability condition for this set of equations yields the monotonic instability threshold which we express in terms of the critical P{\'e}clet number, $\pe_1$. We repeat the solution procedure outlined above for the $n$-th term in the sum~\eqref{asymp_n1_exp} and obtain the following expression for $\pe_1$,
\begin{equation}
  \label{asymp_pe1}
  \pe_1 = 
    \dfrac{2 (2 + 3 \eta) \left( 4 + k_r d^2 \kappa^{1-1/d}   \left( 2 + k_m \kappa \right) \right)}
        {k_m k_r d \, \kappa^2 - 8 D \kappa^{1/d}}
\end{equation}
and $\pe_n$ with $n > 1$,
\begin{multline}
  \label{asymp_pen}
  \pe_n = 4 \left( 2 n + 1 \right) (1 + \eta) 
    \\ \times \left( n \left( n + 1 \right)^2 
      + k_r d^2 \kappa^{1-1/d} \left( n + 1 + k_m \kappa \right) \right)
    \\ / \left( k_m k_r d \, \kappa^2 - 4 D \kappa^{1/d} n \left( n + 1 \right)^2 \right).
\end{multline}
It is easy to see that for any values of the problem parameters, $\pe_1$ remains the lowest and, thus, most ``dangerous'' instability threshold. Recall that $\pe_1$ corresponds to the onset of the mode $\psi_{o,1}$ that implements self-propulsion of the drop. In what follows, we study the flow regimes emerging in vicinity of $\pe_1$, that is, near the onset of spontaneous droplet self-propulsion.

\subsection{Weakly nonlinear analysis (terms at $\epsilon^2$)}
\label{asymp2}
We now proceed with the analysis of the terms at $\epsilon^2$ in the expansion of the dimensionless problem~\eqref{prob_nondim_beg}-\eqref{prob_nondim_end}. Our goal is to study droplet behavior in vicinity of the monotonic instability threshold $\pe_1$. To this end, we also expand the P{\'e}clet number as,
\begin{equation}
  \label{asymp_delta_def}
  \pe = \pe_1 + \epsilon \delta.
\end{equation}

By definition, terms at $\epsilon^2$ should include quadratic interactions of the linear solutions obtained in Sec.~\ref{asymp1}. In particular, general solution presented in Eqs.~\eqref{asymp_n10_sol}-\eqref{asymp_n11_sol} includes the modes proportional to $L_0$ and $L_1$ and can be considered as a spectrum of monotonic perturbations at $\pe = \pe_1$. Properties of Legendre polynomials dictate that quadratic interactions of these modes should include $L_0$, $L_1$, and $L_2$. Specific near and far field solutions may be written as,
%adapted from Refs.~\cite{Morozov19a,Morozov19b},
\begin{multline}
  N^{(2)} (\textbf{r})
  = c_0^{(2)} + \frac{b_0^{(2)}}{r}
  - \frac{\pe_1 a_1^{(1)} N_1^{(1)}}{12 r^4}
  \\ + \frac{\kappa B}{120 r^5} \left( 2 + 5 r^3 + 20 r^6 \right)
  \\ + L_1(\mu) \left( 
      c_1^{(2)} r + \frac{b_1^{(2)}}{r^2}
      - \frac{1 + 2 r^3}{4 r^3} \left( \pe_1 a_1^{(1)} N_0^{(1)} + \kappa A \right)
    \right) \\
  + L_2(\mu) \Bigg(
      c_2^{(2)} r^2 + \frac{b_2^{(2)}}{r^3} - \frac{1 + 2 r^3}{6 r^4} \pe_1 a_1^{(1)} N_1^{(1)}
      \\ + \frac{B \kappa}{168 r^5} \left( 5 + 35 r^3 + 14 r^6 \right)
      - \frac{2 + 3 r^2}{2 r^4} \pe_1 \kappa a_2^{(2)}
    \Bigg),
\end{multline}
\begin{multline}
  \label{asymp_f2_sol}
  F^{(2)}(\boldsymbol \rho) = \bigg( -\frac{\kappa A ( 1 + \mu )}{2}
    + \frac{\pe_1 k_m \kappa^2}{2 \rho ( 1 + k_m \kappa )} \bigg) 
      \\ \times e^{-\pe_1 a_1^{(1)} \rho \left( 1 + \mu \right) / 2},
\end{multline}
where $A \equiv \delta a_1^{(1)} + \pe_1 a_1^{(2)}$, $B \equiv \left( \pe_1 a_1^{(1)} \right)^2$, whereas constant coefficients $c_n^{(2)}$ and $b_n^{(2)}$ are to be determined from matching and boundary conditions. Following the algorithm developed in Ref.~\citenum{Morozov19a}, we match $N^{(2)}(\textbf{r})$ and $F^{(2)}(\boldsymbol \rho)$ and substitute the result into the boundary conditions at $\epsilon^2$ to obtain a set of algebraic equations for the coefficients $a_1^{(1)}$, $a_n^{(2)}$, and $b_n^{(2)}$. Solvability condition of this set of equations in the limit of ${D \rightarrow 0}$ reads,
\begin{equation}
  \label{asymp_sol_cond}
  a_1^{(1)} \left( a_1^{(1)} - a_s \right) = 0
\end{equation}
with
\begin{multline}
  a_s \equiv \delta \pe_1^{-1} ( 1 + k_m \kappa )
  \\ / \left( \pe_1 + \kappa^{-1/d} \left( 2 + 3 \eta \right) 
      \left( 2 \left( d - 1 \right) + k_m \kappa \left( 2 d - 1 \right) \right) \right).
\end{multline}
It is easy to see that solvability condition~\eqref{asymp_sol_cond} admits two solutions. The first, $a_1^{(1)} = 0$, is trivial and corresponds to the motionless base state~\eqref{prob_base_state}. The second describes a droplet that self-propels above the instability threshold (${\delta > 0}$) with a steady velocity.

%%%%%%%%%%%%%%%%%%%
% Numerical method
%%%%%%%%%%%%%%%%%%%
\section{Numerical method}
\label{method}
Numerical method employed in this paper is based on the spectral method used in Ref.~\citenum{Morozov19b}. In particular, we approximate the concentration of swollen micelles in the bulk and concentration of the adsorbed monomers using a truncated series of Legendre harmonics,
\begin{align}
  \label{num_exp_m}
  M(t,r,\mu) &= \sum\limits_{n=0}^{N_\text{mod}} M_n(t,r) L_n(\mu), \\
  \label{num_exp_gam}
  \Gamma(t,\mu) &= \sum\limits_{n=0}^{N_\text{mod}} \Gamma_n(t) L_n(\mu).
\end{align}
\blu{Note that the value of $N_\text{mod}$ determines the angular resolution of the numerical scheme. In particular, this resolution limits the degree of nonlinearity of the chemical reaction~\eqref{prob_m_react} that may be resolved numerically. As a result, micelles with $d \leq 50$ were considered in this paper.}

We substitute expansions~\eqref{num_exp_m}-\eqref{num_exp_gam} along with truncated expansion for the flow field~\eqref{prob_psi_i}-\eqref{prob_psi_o} into dimensionless bulk~\eqref{prob_nondim_m_trans} and surface~\eqref{prob_nondim_end} advection-diffusion equations and obtain a set of $2n$ evolution equations for each of the component of the Legendre spectrum,
\begin{align}
  \label{method_m}
  \partial_t M_n + & \frac{2 n + 1}{2 r^2} \nonumber  
    \\ & \times \sum\limits_{i=1}^{N_\text{mod}} \sum\limits_{j=0}^{N_\text{mod}} \nonumber
      a_i \big( i ( i + 1 ) I_{ijn} \phi_i \partial_r M_j 
        + J_{ijn} M_j \partial_r \phi_i \big) 
    \\ & = \frac{1}{\pe} \left( \partial_{rr} + \frac{2 \partial_r}{r} - \frac{n ( n + 1 )}{r^2} \right) M_n, \\
  \label{method_gamma}
  \partial_t \Gamma_n + & \frac{2 n + 1}{2} \nonumber
    \\ & \times \sum\limits_{i=1}^{N_\text{mod}} \sum\limits_{j=0}^{N_\text{mod}} \nonumber
      \big( - i ( i + 1 ) I_{ijn} + J_{ijn} \big) a_i \Gamma_j \partial_r \phi_i
    \\ & = \frac{1}{\pe} \left( - n ( n + 1 ) \Gamma_n + \rho_n \right)
  \quad \text{at  } r = 1,
\end{align}
where
\begin{align}
  & \phi_n = \begin{cases}
    \left( 1 - r^3 \right) / \left( 2 r \right) \; & n = 1\\
    \left( 1 - r^2 \right) / r^n \; & n > 1
  \end{cases}, \\
  & I_{ijn} = \int\limits_{-1}^1 d\mu L_i L_j L_n, \quad
    J_{ijn} = \int\limits_{-1}^1 d\mu \left( 1 - \mu^2 \right) L_i' L_j' L_n, \\
  & \rho_n = \frac{2n+1}{2} \int\limits_{-1}^1 d\mu \left( e^{-k_m M} - d \, k_r \Gamma^d \right) L_n
    \quad \text{at  } r = 1.
\end{align}

To obtain the values of $a_i$ appearing in Eqs.~\eqref{method_m}-\eqref{method_gamma}, we substitute expansion~\eqref{num_exp_gam} into boundary condition for tangential stresses and obtain,
\begin{equation}
  a_n(t) = \begin{cases}
      2 / \left( 3 ( 2 + 3 \eta ) \right) & n = 1 \\
      1 / \left( 2 ( 2 n + 1 ) ( 1 + \eta ) \right) & n > 1.
    \end{cases}
\end{equation}

Two numerical techniques are used in this paper: time marching and natural continuation. Our time-marching scheme is identical to the one described in Ref.~\citenum{Morozov19b}. Specifically, we introduce an exponentially stretched spatial grid in $r$ and treat nonlinear terms of Eqs.~\eqref{method_m}-\eqref{method_gamma} explicitly, while Crank-Nicholson scheme is used to integrate linear terms. Steady solutions reached in the course of the time-marching are employed to initialize the method of natural continuation. In particular, we refine the steady solution using Newton's iterations, and then use the result as an initial guess for the numerical solution with slightly different values of problem parameters. We repeat the procedure to scan the parameter space of the problem. \blu{To generate the stability maps shown in Fig.~\ref{figMap}, we assessed the linear stability of the resulting steady solutions numerically. To this end, eigenvalues of the Jacobian of the model equations were computed. Steady solution was considered stable when the Jacobian only had eigenvalues with negative real parts.}

To validate the results of our simulations, we repeat the computations for $N_\text{mod} = 20$ and $80$ grid points in $r$; and $N_\text{mod} = 30$ and $120$ grid points in $r$. In Fig.~\ref{validation} we also compare our numerical results with the predictions of asymptotic analysis developed in~Appendix~\ref{asymptotic}.
\begin{figure}[H]
  \centering
  \includegraphics[scale=0.5]{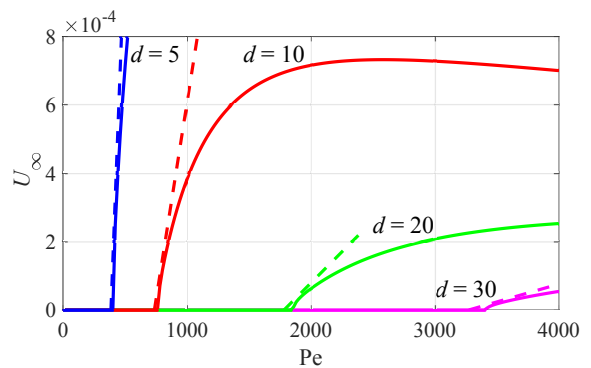}
  \caption{
    Comparison of the results of numerical simulations (solid lines) 
    with the predictions of asymptotic analysis, namely, Eq.~\eqref{asymp_sol_cond} (dashed lines)
    for $D = 10^{-3}$, $\eta = 1$, $k_m = 10$, $k_r = 1$, and various swollen micelle sizes. 
  }
  \label{validation}
\end{figure}

%\balance

%
% Bibliography
%
\bibliographystyle{unsrt}
\bibliography{drop_cmc}

\end{document}